\documentclass[12pt]{article}
\usepackage[latin1]{inputenc}
\usepackage{amsmath,color,graphicx}
\usepackage{amsfonts}
\usepackage{amssymb,amsthm}
\definecolor{Red}{cmyk}{0,1,1,0}

\definecolor{Blue}{cmyk}{1,1,0,0,}

\newcommand{\ba}{\begin{array}}
\newcommand{\ea}{\end{array}}
\newcommand{\be}{\begin{equation}}
\newcommand{\ee}{\end{equation}}
\newcommand{\ben}{\begin{enumerate}}
\newcommand{\een}{\end{enumerate}}

\let\a=\alpha
\let\b=\beta
\let\d=\delta
\let\e=\varepsilon

\let\g=\gamma

\let\l=\lambda

\let\o=\omega

\let\s=\sigma
\let\t=\tau

\let\O=\Omega

\newcommand{\eop}{\nopagebreak\hfill\fbox{ }}

\newcommand{\La}{\mathbb{L}}

\newcommand{\B}{{\mathbb B}}

\newcommand{\Z}{\mathbb{Z}}

\newcommand{\N}{\mathbb{N}}
\newcommand{\T}{\mathbb{T}}

\newcommand{\ufrac}[2]{\genfrac{}{}{0pt}{3}{#1}{#2}}
\newtheorem*{teorema}{Theorem}

\begin{document}
\title{
Decay Properties of the Connectivity  for Mixed Long Range
Percolation Models on $\Z^d$}
\author{
Gast\~ao A. Braga, Leandro M. Ciolleti and Rémy Sanchis
\\
\small{Departamento de Matem\'atica - UFMG}\\
\small{Caixa Postal 1621} \small{30161-970 - Belo Horizonte - MG -
Brazil} } \maketitle
\begin{abstract}
In this short note we consider mixed short-long range
independent bond percolation models on $\Z^{k+d}$.
Let $p_{uv}$ be the probability that the edge $(u,v)$
will be open. Allowing a $x,y$-dependent length scale
and using a multi-scale analysis due to Aizenman and Newman, we show that the long
distance behavior of the connectivity $\tau_{xy}$ is governed by the
probability $p_{xy}$. The result holds up to the critical point.
\end{abstract}
\section{Introduction}
\label{sec:intro} In this short note we consider a long range
percolation model on $\La = (\Z^{k + d}, \B)$, where $u \in \Z^{k+d}$
is of the form $u = (\vec u_0,\vec u_1)$, with  $\vec u_0 \in \Z^k$ and
$\vec u_1 \in \Z^d$ and $\B$ is the set
of edges (unordered pairs) $(u,v)$, $u\not= v \in \Z^{k+d}$. To each
edge $(u,v)$ we associate a Bernoulli random variable $\omega_b$ which is open ($\o_b=1$)
with probability
\be
\label{eq:bond-prob} p_{uv} = p_{uv}(\b) \equiv \b J_{uv},
\,\,\,\,\,\,\, u, v\in\Z^{k + d} \ee where $\b \in [0,1]$ and, for
$\epsilon > 0$, $J_{uv}$ is
 \be
\label{modified}
 J_{uv}=\begin{cases}2(1+\|\vec u_1-\vec v_1 \|^{d+\e})^{-1}&
\mbox{ if } \vec u_0 =\vec v_0 \mbox{ and } \vec u_1 \not= \vec
v_1 ;\\ 1 &\mbox{ if } \vec u_1 =\vec v_1 \mbox{ and } \| \vec u_0
-\vec v_0 \|=1;\\ 0 & \mbox{ otherwise}.
\end{cases}
\ee
We denote the event $\{\o\in\O : \mbox{there is an open path connecting
$x$ to $y$} \}$ by $\{x\leftrightarrow y\}$ and define the connectivity function
by $\tau_{xy}\equiv P\{x\leftrightarrow y\}$.
Let $\| x \|=|x_1|+\cdots +|x_d|$ be
the $L^1$ norm on $\Z^d$ and $\b_c = \mbox{sup}\{\b\in[0,1] :
\chi(\b)<\infty\}$. Our aim is to show the following result
\begin{teorema}
\label{teo:dec-misto} Suppose $\b<\b_c$ and consider the long range
percolation model with $p_{uv}$ given by (\ref{eq:bond-prob}) and
$J_{uv}$ given by (\ref{modified}). Then there exist positive
constants $C=C(\b)$ and $m=m(\b)$ such that \be \label{eq:epdcp}
\tau_{xy}\leq \frac{C\ \ e^{-m\|\vec x_0-\vec
y_0\|}}{1+\|\vec{x}_1-\vec{y}_1\|^{d+\e}} \ee
for all $x,y\in \Z^{k+d}$.
\end{teorema}

The above result says that que probability $p_{uv}$ dictates the
long distance behavior of the connectivity function in the
subcritical regime (similar lower bounds are easily obtained from
the FKG inequality).  For the one dimensional $\Z^{0+1}$
percolation model, the above result is known to hold, see
\cite{bib:aize-newm}, the same being true for one dimensional
$\Z^{0+1}$ $O(N)$ spin models, $1\leq N \leq 4$, see
\cite{bib:spohn-zwerger}. The result is expected to hold in the
$d$-dimensional $\Z^{0+d}$ lattice but it is not clear how to
prove it if $\b < \b_c$, although one can see it holds if $\b
\approx 0$. Our upper bound (\ref{eq:epdcp}) holds if $\b < \b_c$
and for $(k+d)$-dimensional lattices, $k\geq 0$ and $d\geq 1$.
For lattice spin models, the upper bound (\ref{eq:epdcp}) is known
at the high temperature regime, see Ref. \cite{bib:gross} for
bounded spin models and Ref. \cite{bib:isra-napp} for unbounded
(and discrete) ones. Ref. \cite{bib:aldo-bene} extends some of the
results of \cite{bib:gross, bib:isra-napp} to a general class of
continuous spin systems, with $J_{uv}$ given by
(\ref{eq:bond-prob}) and $u, v \in \Z^{0+d}$, while
\cite{bib:rafa-emma-aldo} considers the more general mixed decay
model. In both cases, the polymer expansion (see
\cite{bib:glimm-jaffe} and references therein) is used and the
results hold only in the perturbative regime.

The Hammersley-Simon-Lieb inequality
\cite{bib:hammersley,bib:Simon-ineq,bib:lieb}
is a key ingredient in \cite{bib:aize-newm} and \cite{bib:spohn-zwerger} and
here we also adopt this ``correlation inequality"
point of view. For completeness, we state this inequality
in the form we will use, see \cite{bib:aize-newm2}. For each set
$S\subset\Z^d$, let $\tau^{S}_{xy}\equiv P\{x\leftrightarrow y\
\mbox{inside}\ S\}$. Then

{\bf Hammersley-Simon-Lieb Inequality (HSL)}{\it
\label{teo:simo-lieb} Given $x, y \in \Z^{k+d}$, if $S\subset\Z^{k+d}$ is
such that $x\in S$ and $y\in S^c$, then}
$$
\tau_{xy} \leq \sum_{\{ u\in S, v\in S^c\}} \tau^{S}_{xu}\,\,
p_{uv}\,\, \tau_{vy}.
$$

We now recall some known facts about the long range
percolation model defined by (\ref{eq:bond-prob})  and
(\ref{modified}). Let $\theta(\b , \e) = P_{\b,
\e}\{0\leftrightarrow \infty\}$ be the probability that the origin
will be connected to infinity. If $k+d\geq 2$ is the space
dimension then, by comparing with the nearest neighbor model and
for any positive $\e$, there exists  $\b_c = \b_c(d, \e)$ such
that $\theta(\b, \e)=0$ if $\b< \b_c$ and $\theta(\b, \e)>0$ if
$\b> \b_c$. For $k=0$ and $d=1$, it is known that the existence of
$\b_c$ depends upon $\e$, if $\e >1$ then there is no phase
transition \cite{bib:schulman} while it shows up if $0\leq \e \leq 1$,
see \cite{bib:newm-schu}. A phase transition can also be measured in
terms of $\chi$, the mean cluster size, given by $\chi = \sum_x
\t_{0x}$. Let $\pi_c(d, \e)= \mbox{sup}\{\b: \chi(\b,
\e)<\infty\}$. Then, it comes from the FKG Inequality
\cite{bib:grimmett} that $\pi_c(d, \e)\leq \b_c(d, \e)$. The
equality $\pi_c(d, \e)= \b_c(d, \e)$ holds for the class of models
we are dealing with and it was proved independently by Aizenman and Barsky  in
\cite{bib:aize-bars} and Menshikov in \cite{bib:menshikov}. We will use the
condition $\chi<\infty$ to characterize the subcritical region.

The remaining of this note is divided as follows: in the next section we
prove Theorem \ref{teo:dec-misto} and in Section \ref{sec:conc-rema} we make
some concluding remarks regarding the validity of our results to
ferromagnetic spin models.


\section{Proof of the Theorem}
\label{sec:dtcp}

Let $x=(\vec x_0,\vec x_1)\in \Z^{k+d}$. We first observe that
$$
\chi=\sum_{n\geq 0}\ \sum_{\|\vec x_0\|=n}\
\sum_{\vec{x}_1\in\Z^d}\tau_{0x}.
$$
Since $\chi<\infty$, given $\l\in(0,1)$, there exists
$n_0\in \N$ such that, for all $n\geq n_0$,  we have
$$
\sum_{\|\vec x_0\|=n}\ \sum_{\vec{x}_1\in\Z^d}\tau_{0x}<\l.
$$
Consider now $x=(\vec x_0,\vec x_1)$ with $\|\vec x_0\|>n_0$. Using the
translation invariance of the model and applying iteratively the HSL
Inequality with $S= \{x\in \Z^{k+d};\|\vec x_0\|\leq n_0\}$, we obtain
$$
\sum_{\vec x_1 \in\Z^d}\t_{0x}\leq \l^{\lfloor \|\vec x_0\|/n_o  \rfloor} \leq
C_1\exp(-(m+\d)\|\vec x_0\|),
$$
where $\lfloor r  \rfloor$ denotes the integer
part of $r$,  $\delta > 0$ is given and $m$ is defined by
 $e^{-(m+\d)} = \lambda^{1/n_0}$.
Next we show that
a HSL type inequality holds for the modified connectivity function
$T_{m}(x,y)\equiv e^{m\|\vec{x}_0-\vec{y}_0\|}\tau_{xy}$ and for
the set $S= \mathcal{C}_r(x)\equiv \{z\in\Z^{k+d};\|\vec{x}_1-\vec{z}_1\|\leq r\}$.
Applying the HSL Inequality with the above specified $S$, we have
$$
\t_{xy}\leq \sum_{\ufrac{u\in\mathcal{C}_r(x)}{v\in\mathcal{C}^c_r(x)} }\t_{xu}p_{uv}\t_{vy}.
$$
Then, for $y\in\mathcal{C}^c_L(x)$ for some $L>1$, we obtain
$$
\begin{array}{rcl}
 T_{m}(x,y)&=&e^{m\|\vec x_0-\vec y_0\|}\tau_{xy}
\leq e^{m\|\vec x_0-\vec y_0\|}\displaystyle\sum_{{\ufrac{u\in\mathcal{C}_L(x)}{
v\in\mathcal{C}^c_L(x)}}} \tau_{xu}p_{uv}\tau_{vy}\\[1cm]&\leq&
\displaystyle\sum_{{\ufrac{u\in\mathcal{C}_L(x)}{
v\in\mathcal{C}^c_L(x)}}}e^{m\|\vec x_0-\vec u_0\|}\tau_{xu}\ \ p_{uv}\ \
e^{m\|\vec y_0-\vec v_0\|}\tau_{vy}
\end{array}
$$
since we necessarily have that $\vec u_0=\vec v_0$. It then follows that
$$
T_{m}(x,y)\leq \sum_{{\ufrac{u\in
\mathcal{C}_L(x)}{ v\in\mathcal{C}^c_L(x)}}} T_{m}(x,u)p_{uv}T_{m}(v,y).
$$
We remark that
$
\chi_{m}\equiv \sum_{x\in \Z^{k+d}}T_{m}(0,x)<\infty
$
is finite if $\b < \b_c$ since
$$
\begin{array}{rcl}
\displaystyle\sum_{x\in \Z^{k+d}}T_{m}(0,x)=\sum_{x\in
\Z^{k+d}}e^{m\|\vec x_0\|}\tau_{0x}&\leq&\displaystyle\sum_{k\geq
0}\displaystyle\sum_{\|\vec x_0\|=k}e^{m\|\vec x_0\|}\sum_{\vec{x}_1\in\Z^d}\tau_{0x}\\[0.6cm]
&\leq&\displaystyle\sum_{k\geq 0}2dk^{d-1} e^{-\d k} < \infty.\\[0.6cm]
\end{array}
$$
From now on we closely follow Section 3 of \cite{bib:aize-newm} and prove the
polynomial decay of  $T_{m}$ up to the
critical point.
For fixed $x, y\in\Z^{k+d}$ and  $L\equiv
\|\vec{x}_1-\vec{y}_1\|/4$, we know that
\begin{eqnarray}
\label{eq:bound-tau}
T_{m}(x,y)&\leq& \displaystyle\sum_{{\ufrac{u\in
\mathcal{C}_L(x)}{ v\in\mathcal{C}^c_L(x)}}}
T_{m}(x,u)p_{uv}T_{m}(v,y)\nonumber\\[0.3cm]
&\leq& \displaystyle\!\!\!\!\!\!\!\!\!\sum_{{\ufrac{u\in
\mathcal{C}_L(x)}{v\in\mathcal{C}^c_L(x)\cap\mathcal{C}_{3L}(x)}}}\!\!\!
T_{m}(u,x)p_{uv}T_{m}(v,y) + \displaystyle\!\!\!\!\!\!\sum_{{\ufrac{u\in
\mathcal{C}_L(x)}{v\in\mathcal{C}^c_L(x)\cap\mathcal{C}^c_{3L}(x)}}}\!\!\!T_{m}(x,u)p_{uv}T_{m}(v,y).
\end{eqnarray}
Let
$$
\T_{m}(L)\equiv \sup\{T_{m}(0,u); u\in \mathcal{C}^c_L(0)\}
\,\,\,\,\,\,
\mbox{and}
\,\,\,\,\,\,
\g_L\equiv\displaystyle\sum_{{\ufrac{u\in\mathcal{C}_L(x)}{
v\in\mathcal{C}^c_L(x)}}} T_{m}(x,u)p_{uv}.
$$
Then, the first term on the r.h.s. of (\ref{eq:bound-tau}) is
bounded above by $\T_{m}(L/2)\g_L$ while the second one is bounded by
$$
\frac{2^{d+\e}  2 \b  \chi_{m}^2}{1+\|\vec{x}_1-\vec{y}_1\|^{d+\e}},
$$
leading to
$$
T_{m}(x,y)\leq \frac{2^{d+\e}  2 \b
\chi_{m}^2}{1+\|\vec{x}_1-\vec{y}_1\|^{d+\e}}+
\g_L\T_{m}\left(\frac{L}{2}\right).
$$
Now, since $\chi_{m}<\infty$ for $\b < \b_c$ and since
$\sum_up_{0u}<\infty$, we have that $\g_L\to 0$ as $L\to\infty$.
For $\a\in(0,2^{-(d+\e)})$, there exists $L_0>0$ such that
$\g_{L}<\a$ for all $L\geq L_0$. Considering $L>L_0$, it follows
that
\be \label{ineq:L-2} \T_{m}(L)\leq \frac{2^{d+\e}  2 \b
\chi_{m}^2}{1+\|\vec{x}_1-\vec{y}_1\|^{d+\e}}+\a\T_{m}\Big(\frac{L}{2}\Big).
\ee
Iterating (\ref{ineq:L-2}) $n$ times, with $n$ the smallest
integer for which $L2^{-n}\leq L_0$, we have for all $L>L_0$
$$
\T_{m}(L)\leq \frac{ 2 \b \chi_{m}^2\sum_{j=0}^{n-1}(\a
2^{d+\e})^j}{1+\|\vec x_1-\vec y_1\|^{d+\e}} +\a^n \T_{m}\Big(\frac{L}{2^n}\Big).
$$
Noting that $T_{m}(x,y)\leq \T_{m}(L)$, that $\T_{m}(L)\leq 1$ for any $L >
0$ and that
$$
\a^n\leq
2^{-(d+\e)n}=\frac{1}{(1+L^{d+\e})}\frac{(1+L^{d+\e})}{2^{(d+\e)n}}
\leq
\frac{2\cdot 2^{d+\e}}{1+\|\vec{x}_1-\vec{y}_1\|^{d+\e}}\left(\frac{L}{2^n}\right)^{d+\e}\leq
\frac{2(2L_0)^{d+\e}}{1+\|\vec{x}_1-\vec{y}_1\|^{d+\e}},$$
we can conclude that, for $\b < \b_c$,
$$
\tau_{xy}\leq \frac{C\ \ e^{-m\|\vec x_0-\vec
y_0\|}}{1+\|\vec{x}_1-\vec{y}_1\|^{d+\e}}
$$
and the bound (\ref{eq:epdcp}) holds.
\eop

\section{Concluding Remarks}
\label{sec:conc-rema}

The strategy used to prove the main Theorem can also
be applied to  $\Z^d$ ferromagnetic models with free boundary
conditions and pair interaction $J_{uv}$
given by (\ref{modified}). For this class of models the
Griffiths inequalities \cite{bib:griffiths-I, bib:griffiths-II}
are valid, guaranteeing the positivity of spin-spin correlations
$\langle \s_x\s_y \rangle$. Simon-Lieb Inequality
 also holds in this case (see
\cite{bib:glimm-jaffe} and references therein), with $p_{uv}$
replaced by $\b J_{uv}$, where $\b$ is the inverse of the
temperature, and with $\tau_{xy}$ replaced by $\langle \s_x\s_y
\rangle$. Finally, since the uniqueness of the critical point is
guaranteed in \cite{bib:aize-bars-fern}, the results of Section
\ref{sec:dtcp} are also valid for these models.

{\bf Acknowledgements:} G. B. was partially supported by CNPq;
L. C. acknowledges CNPq for a graduate scholarship;
R. S. was partially supported by Pró-Reitoria
de Pesquisa - UFMG under grant 10023.

\end{document}